\documentclass[fleqn,10pt]{wlscirep}
\title{Three-waveform bidirectional pumping of single electrons with a silicon quantum dot}

\author[1,*]{Tuomo Tanttu}
\author[2,$\dagger$]{Alessandro Rossi}
\author[1]{Kuan Yen Tan}
\author[1]{Akseli M\"akinen}
\author[2]{Kok Wai Chan}
\author[2,+]{Andrew S. Dzurak}
\author[1,+]{Mikko M\"ott\"onen}

\affil[1]{Aalto University, QCD Labs, COMP Centre of Excellence, Department of Applied Physics, Aalto, 00076, Finland}
\affil[2]{University of New South Wales, School of Electrical Engineering \& Telecommunications, Sydney, 2052, Australia}
\affil[*]{tuomo.tanttu@aalto.fi}

\affil[$\dagger$]{present address: Cavendish Laboratory, University of Cambridge, J.J. Thomson Avenue CB3 0HE, Cambridge, U.K.}

\affil[+]{these authors contributed equally to this work}


\begin{abstract}
Semiconductor-based quantum dot single-electron pumps are currently the most promising candidates for the direct realization of the emerging quantum standard of the ampere in the International System of Units. Here, we discuss a silicon quantum dot single-electron pump with radio frequency control over the transparencies of entrance and exit barriers as well as the dot potential. We show that our driving protocol leads to robust bidirectional pumping: one can conveniently reverse the direction of the quantized current by changing only the phase shift of one driving waveform with respect to the others. We also study the improvement in the robustness of the current quantization owing to the introduction of three control voltages in comparison with the two-waveform driving. We anticipate that this pumping technique may be used in the future to perform error counting experiments by pumping the electrons into and out of a reservoir island monitored by a charge sensor.
\end{abstract}
\begin{document}

% control on each barrier as well as on the dot potential

\flushbottom
\maketitle
\thispagestyle{empty}

%\noindent Please note: Abbreviations should be introduced at the first mention in the main text – no abbreviations lists. Suggested structure of main text (not enforced) is provided below.

\section*{Introduction}

After a quarter of a century of development of charge pumps, we are close to redefining the International System of Units (SI) standard for the electrical current, the ampere, such that it would be based on a fixed value of the elementary charge \cite{Pekola2013,Kaestner2015}. The direct experimental realization of such a quantum ampere standard is based on charge pumps which transfer accurately an integer $n$ number of electrons per cycle from the source to the drain at frequency $f$, yielding direct current $I=nef$. The parameter region where the pumped current is quantized in such a way and where it is insensitive to changes in the system parameters is referred to as a plateau. For practical realizations of the current standard and for the closure of the so-called quantum metrology triangle, it is sufficient that the pump yields a current of hundreds of picoamperes with relative accuracy of $10^{-8}$\cite{Feltin2009, Drung2015, Drung2015b}.

%A pump that satisfies these conditions can also be used to close the so-called quantum metrology triangle that relates three different quantum electrical effects and their constants: Josephson effect (Josephson constant, $2e/h$, where $h$ is the Planck constant), quantum Hall effect (von Klitzing constant, $h/e^2$), and single charge pumping (elementary charge, $e$). The closure of the quantum metrology triangle provides a consistency check for the values of these constants \cite{Feltin2009}.

The very first charge pumps were able to produce currents of a few picoamperes with accuracies of a few per cent \cite{Geerligs1990,Geerligs1991,Kouwenhoven1991,Pothier1992}. After this, several different implementations of charge pumps have been proposed and tested: normal-metal tunnel junction devices \cite{Keller1996, Keller1999}, superconducting devices \cite{Niskanen2005, Vartiainen2007, Mottonen2008}, superconductor--normal--metal hybrid turnstiles \cite{Pekola2008, Maisi2009, Kemppinen2009,Kemppinen2009b, Maisi2011, Peltonen2015}, and surface acoustic wave devices \cite{Shilton1996,Talyanskii1997}. At the moment, the most promising candidates for the emerging quantum ampere are semiconductor quantum dots \cite{Blumenthal2007,Jehl2012, Giblin2012, Stein2015, Giblin2016,  Connolly2013, Fujiwara2004, Fujiwara2008, Fujiwara2001, Chan2011, Rossi2014, Jo2015} and single-atom impurities in semiconductors \cite{Yamahata2014,Tettamanzi2014, Lansbergen2012, Roche2012}, the state of the art being, a current of 87 pA in a GaAs/AlGaAs quantum dot with an uncertainty of less than 0.2 parts per million~(ppm) \cite{Stein2015}. Silicon single-electron pumps have also been studied widely\cite{Chan2011,Rossi2014, Fujiwara2001, Ono2003, Yamahata2011, Yamahata2014, Yamahata2014_revB, Yamahata2015, Jo2015, Tettamanzi2014, Hollosy2015}. Benefits of silicon pumps are that they are based on technologies well-known by the industry and they may exhibit suppressed $1/f$ noise and the absence of large random charge jumps\cite{Zimmerman2007, Hourdakis2008, Zimmerman2008, Koppinen2013, Zimmerman2014}.

%Another widely studied platform for the semiconductor single-electron pumps is silicon
%One of the benefits of the silicon pumps is that they are based on metal--oxide--semiconductor (MOS) technology well-known by the semiconductor industry \cite{Rossi2015}. In addition, certain type of devices based on silicon have exhibited suppressed $1/f$ noise and absence of large background charge jumps .

The accuracy of a charge pump can be determined with a charge sensor that monitors the charge state of a reservoir island into which electrons are pumped. Several different charge sensing schemes have been demonstrated: pumping electrons into and out of the reservoir\cite{Keller1996,Kautz1999,Yamahata2011,Yamahata2014_revB}, pumping a number of electrons into the reservoir and cyclically emptying it\cite{Tanttu2015}, monitoring multiple reservoirs interleaved with pumps in a series configuration\cite{Fricke2013,Fricke2014,Peltonen2015}, and monitoring the pump dot without any storage node\cite{Giblin2016}. In general, pumping electrons into and out of a reservoir in semiconductor devices is highly nontrivial due to the asymmetry of the devices and pumping protocols.

In this paper, we demonstrate bidirectional electron pumping in a silicon-based quantum dot by employing a three waveform protocol, thus offering a step towards error counting based on a reservoir dot. Our technology allows simultaneous control over both barriers and the dot potential, enabling convenient switching between the pumping by changing only the phase of one driving signal. This kind of switching between the directions has been demonstrated before in silicon devices\cite{Ono2003, Jehl2012} with two driving waveforms to the barrier gates\cite{Kouwenhoven1991} and in metallic pumps with three waveforms \cite{Pothier1992, Mottonen2008}. We also show the improvement of this pumping process over a two-waveform drive using the same driving amplitudes. This is confirmed by the study of the sensitivity of the current quantization to experimental parameters such as the plunger gate (PL) dc voltage and the rf amplitude of the drive.

\section*{Results}

\subsection*{Pumping with three waveforms}

\begin{figure}[ht]
\centering
\includegraphics[width=130 mm]{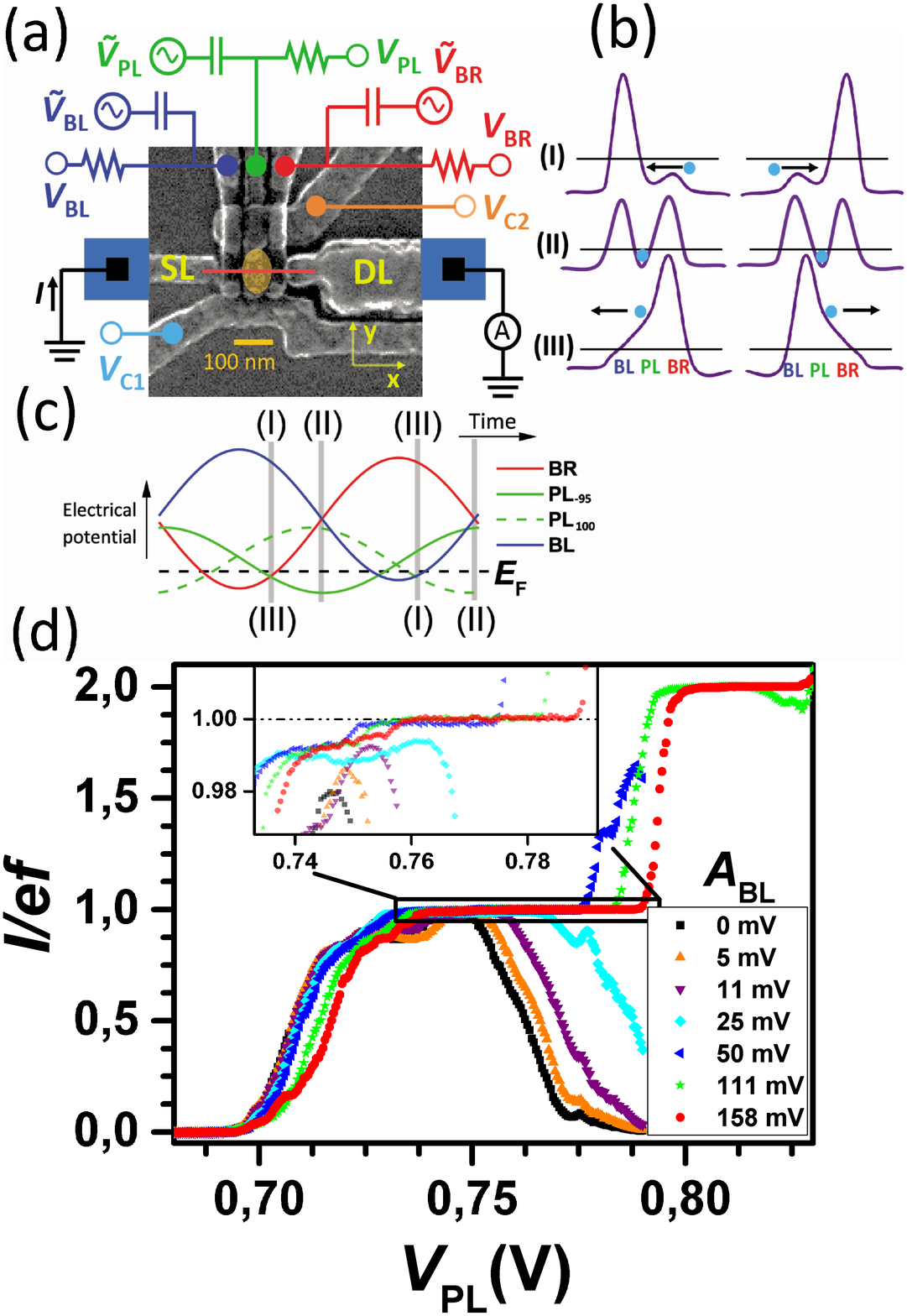}
\caption{Sample and pumping protocol. (a) Scanning electron microscope image of a sample similar to the one used in the experiments together with the schematic measurement setup. The lateral position of the quantum dot is indicated by an orange oval. Blue squares indicate the ohmic contacts of source and drain to the two-dimensional electron gas. (b) Schematic potential landscapes for an electron along the red line in (a) at different stages of the pumping process for two different phases of plunger drive leading to either positive (left) or negative (right) pumped current. The whole process consists of electron loading (I), trapping (II), and unloading (III). (c) Time dependence of the electrical potentials (blue, red and green lines) in the three-waveform driving scheme with two different phases of the plunger drive with respect to BL drive. The Fermi level of the leads is shown by the black dashed line. The gray lines indicate the time instants visualized in (b). (d) Pumped current as a function of plunger gate voltage, $V_{\textrm{PL}}$, with different amplitudes of the driving voltage on the left barrier, $A_\textrm{BL}$, at 200 MHz frequency. Inset: The pumped current from the main panel in the vicinity of plateau $I/ef= 1$. The dc gate voltages defined in (a) assume the following values: $V_{\textrm{SL}}= 3.5$~V, $V_{\textrm{DL}}= 1.8$~V, $V_{\textrm{C1}}= -0.40$~V, $V_{\textrm{C2}}= -0.50$~V, $V_{\textrm{BL}}= 0.80$~V, and $V_{\textrm{BR}}= 0.72$~V. The rf driving amplitudes are $A_{\textrm{BR}}= 158$~mV, and $A_{\textrm{PL}}= 79$~mV with phase difference $\phi_\textrm{BR-PL}= 95^{\circ}$.}
\label{sample}
\end{figure}

A scanning electron microscope image of our quantum dot device and a schematic measurement set-up are presented in Fig.~\ref{sample}(a). The details of forming a two-dimensional electron gas (2DEG) and a single-electron dot isolated from the leads are discussed in Refs.~\citenum{Rossi2014, Rossi2015, Tanttu2015}. Initially, we pump with sinusoidal radio frequency (rf) waveforms applied to PL and to the barrier right gate (BR) with a phase difference of PL with respect to BR, $\phi_{\textrm{BR-PL}}=95^\circ$ at 800 mK temperature. We electrically confine the dot by setting negative bias on the confining gates C1 and C2, with gate voltages $V_\textrm{C1}$ and $V_\textrm{C2}$, respectively\cite{Rossi2014,Seo2014}. Then we search in the dc parameter space for a stable pumping plateau in the positive current direction, i.e., BR corresponds to the entrance barrier and the barrier left gate (BL) to the exit barrier. Here, we define the first plateau as the parameter region for which the normalized pumped current, $I/ef$ is within 5\% of unity. Subsequently, we decrease the dc potential on BL, $V_{\textrm{BL}}$, i.e., we make the exit barrier more opaque, until we measure only a narrow pumping plateau as a function of PL and BR dc voltages, $V_{\textrm{PL}}$ and $V_{\textrm{BR}}$ respectively. At this point, our pumping process is limited by the unloading process. Then we drive BL a sinusoidal waveform 180$^\circ$ phase-shifted with respect to BR so that all the waveforms are sines by $\tilde{V}_{\textrm{BR}} = A_{\textrm{BR}}\sin( \omega t)$, $\tilde{V}_{\textrm{PL}} = A_{\textrm{PL}}\sin( \omega t+\phi_{\textrm{BR-PL}})$, and $\tilde{V}_{\textrm{BL}} = A_{\textrm{BL}}\sin( \omega t+180^{\circ})$. %We control $A_{\textrm{BL}}$ by changing the attenuation of the rf line.
%it'd be better to mention here that the way to form a 2DEG, electrostatically isolate a single electron in the dot etc are described in our previous papers and give references.

A schematic illustration of the pumping process with three pulses is presented in Fig. \ref{sample}(b). First (I) we lower the entrance barrier and the potential of the dot such that an electron tunnels into the dot. In the second phase, (II) the entrance barrier is raised and the electron is trapped in the dot. Then the electron is unloaded (III) by lowering the exit barrier and raising the dot potential. Depending on the phase of PL it is possible that electrons exit at energies above the Fermi level of the leads. Hence it is possible that the electron escapes to the source but such process unlikely due to the high opacity of the entrance barrier. The time dependence of the potentials with two different phases on PL is presented in Fig.~\ref{sample}(c). The normalized pumped current, $I/ef$, at $f=200$~MHz is shown as a function of $V_{\textrm{PL}}$ with varying $A_\textrm{BL}$ in Fig.~\ref{sample}(d). The length of the first plateau increases significantly with higher $A_\textrm{BL}$. Note also that with low amplitudes there is no second plateau, but with high amplitudes the second plateau corresponding to transfer of two electrons per cycle is clearly visible.

Not only does the length of the plateaus increase with increasing $A_\textrm{BL}$ but the current quantization becomes more accurate, i.e., the pumped current is closer to the expected value. The inset of Fig.~\ref{sample}(d) shows that with low amplitude, the normalized pumped current at the first plateau is below unity by a few per cent. However, with high amplitudes the normalized current reaches unity more precisely which indicates a more robust unloading process. 

%Furthermore, we note that the length of this accurate region in $V_\textrm{PL}$ also increases as a function of $A_\textrm{BL}$.

\subsection*{Stability of the pumping process}

\begin{figure}[ht]
\centering
\includegraphics[width=\linewidth]{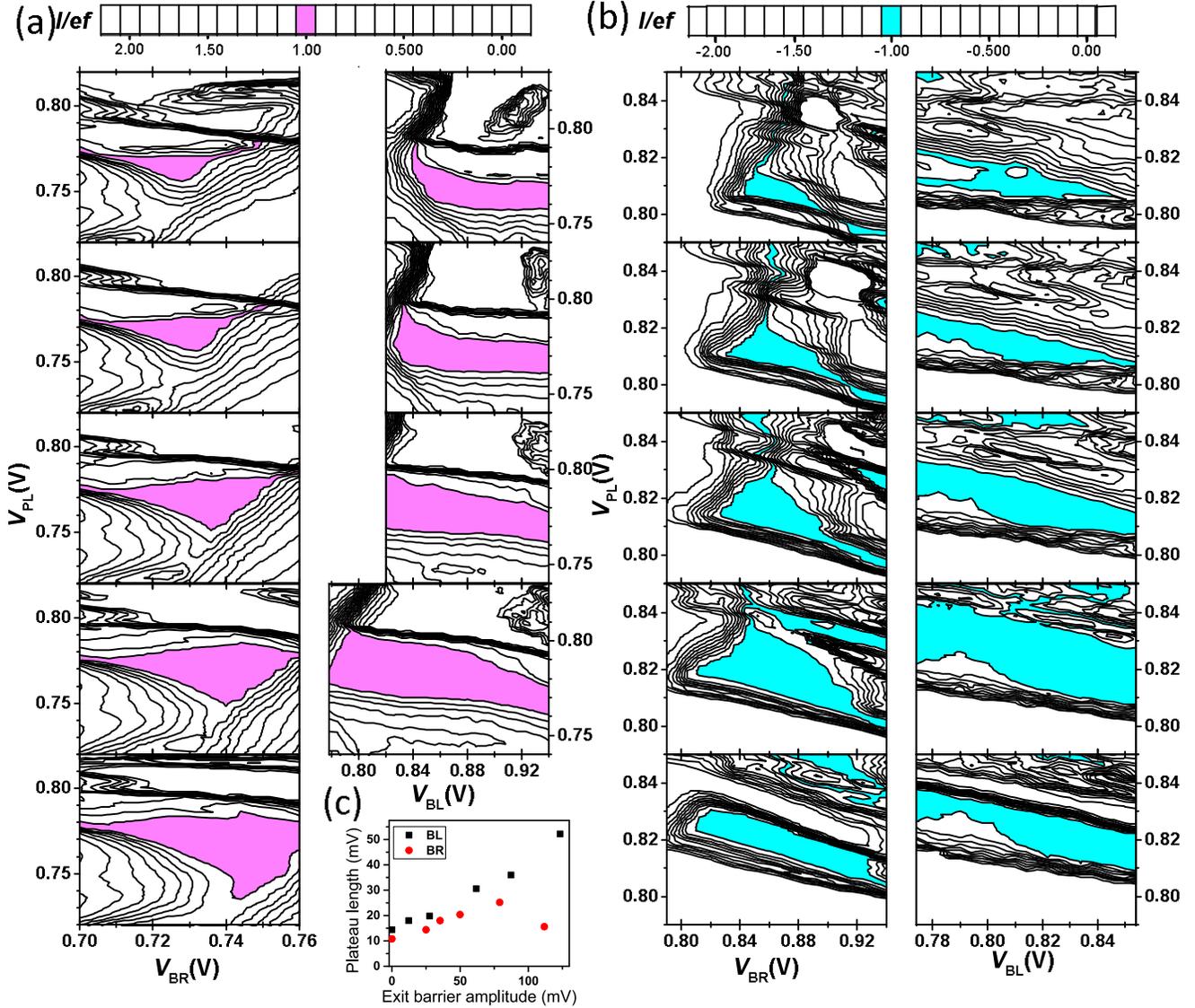}
\caption{Stability of the pumped current. Pumped current as a function of the plunger gate voltage, $V_{\textrm{PL}}$, and the barrier right gate voltage, $V_{\textrm{BR}}$, (left columns) and of the plunger gate voltage and the barrier left gate voltage, $V_{\textrm{BL}}$, (right columns) with different amplitudes on BL (a) and BR (b). The frequencies are 200 MHz in (a) and 100 MHz in (b). Magenta (a) and cyan (b) color indicate pumping regions where $I/ef=\pm 1$ within 5\%. The parameter values used in (a) are the same as in Fig.~\ref{sample}(d) except $V_{\textrm{BL}}= 0.9$~V in left column and $V_{\textrm{BR}}= 0.725$~V in the right column and $A_{\textrm{BR}}=123 $~mV in both columns. In (b) the values are the same as in (a) except $V_{\textrm{BL}}= 0.814$~V in left column and $V_{\textrm{BR}}= 0.85$~V in the right column, $A_{\textrm{BL}}= 112$~mV and $A_{\textrm{PL}}= 205$~mV in both columns and $\phi_\textrm{BL-PL}= 120^{\circ}$. (c) Maximum plateau length in the plunger dc voltage as a function of exit barrier amplitude ($A_\textrm{PL}$ or $A_\textrm{BR}$) extracted from (a) (black squares) and (b) (red circles). The parameter values are the same as in (a) and~(b) for the BL and BR, respectively. }
\label{comparison}
\end{figure}

We also study the stability of the pumping process by measuring the pumped current as a function of $V_{\textrm{PL}}$, $V_{\textrm{BR}}$, and $V_{\textrm{BL}}$. We begin by pumping with PL and BR as described above, but decrease $A_{\textrm{BR}}$ until we measure only a narrow plateau region in terms of $V_{\textrm{PL}}$, $V_{\textrm{BR}}$, and $V_{\textrm{BL}}$. We measure the pumped current as a function of $V_{\textrm{PL}}$ and $V_{\textrm{BR}}$ and as a function of $V_{\textrm{PL}}$ and $V_{\textrm{BL}}$. We repeat these scans with different $A_\textrm{BL}$ values. The results are shown in Fig.~\ref{comparison}(a). We observe that the plateau enlarges in the parameter space ($V_{\textrm{PL}}$, $V_{\textrm{BR}}$, and $V_{\textrm{BL}}$) with increasing $A_\textrm{BL}$. The maximum length of the plateau in terms of $V_\textrm{PL}$ increases from 14 to 52 mV as $A_\textrm{BL}$ increases from 0 to 123 mV.

This experiment is repeated in the case where the exit barrier is BR, i.e., the opposite pumping direction. We pump initially with PL and BL using PL phase difference w.r.t. BL as $\phi_\textrm{BL-PL}=120^\circ$ at 100 MHz. A third waveform (complement waveform to that of BL) is applied to BR. The waveforms are: $\tilde{V}_{\textrm{BL}} = A_{\textrm{BL}}\sin (\omega t)$, $\tilde{V}_{\textrm{PL}} = A_{\textrm{PL}}\sin (\omega t+\phi_{\textrm{BL-PL}})$, and $\tilde{V}_{\textrm{BR}} = A_{\textrm{BR}}\sin (\omega t+180^{\circ})$. We observe a similar widening of the plateau in Fig.~\ref{comparison}(b) as in Fig.~\ref{comparison}(a). However, in this case the length of the plateau in $V_\textrm{PL}$ reaches a maximum around $A_\textrm{BR}=79$~mV, inferring that the BR gate couples to the quantum dot more strongly than the BL gate. For simplicity, we do not compensate this coupling in the experiment. Therefore the increased amplitude in BR interferes with the pumping process more than that in BL, restricting our ability to improve the pumping process beyond the observed optimal point. The maximum length of the plateau in terms of $V_\textrm{PL}$ increases from 12 mV to 25 mV as $A_\textrm{BR}$ increases from 0 to 79 mV. % The maximum plateau length does not reach as high value as in the previous experiment. This is most likely due to the high coupling of BR to the dot and due to the fact that we are using higher amplitude on PL. 

The maximum length of the plateau in $V_\textrm{PL}$ as a function of the exit barrier amplitude extracted from the data of Fig.~\ref{comparison}(a) and~\ref{comparison}(b) is presented in Fig.~\ref{comparison}(c). In both scenarios, the length increases rather linearly as we increase the amplitude of the exit barrier, but in the case of BR, we observe a maximum.

\subsection*{Bidirectional pumping}

\begin{figure}[ht]
\centering
\includegraphics[width=\linewidth]{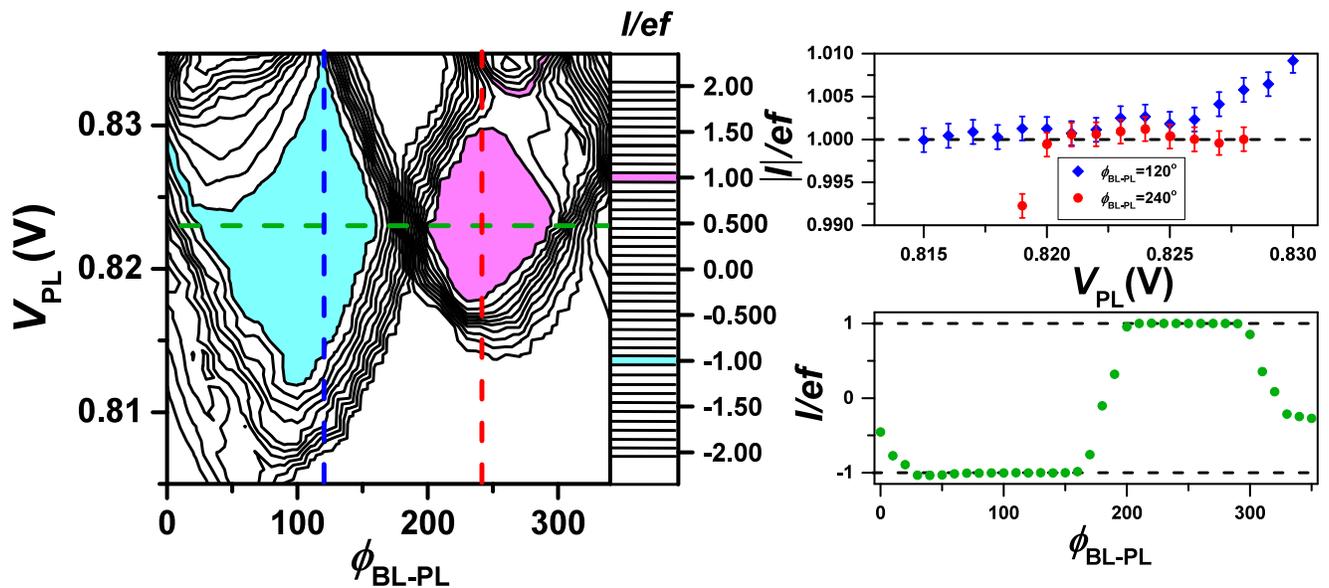}
\caption{Bidirectional pumping. (a) Pumped direct current as a function of the plunger dc voltage and the phase difference of plunger rf driving voltage with respect to the barrier left rf drive. Magenta and cyan color indicate the regions where $I/ef=\pm 1$ within 5\%, i.e., where one electron is pumped per cycle in the positive and negative directions, respectively. The parameter values are the same as in Fig.~\ref{comparison}(b) except $V_\textrm{BL}=0.814$~V, $V_\textrm{BR}=0.850$~V, and $A_\textrm{BR}=79$~mV. (b) Absolute values of the cross sections along $V_\textrm{PL}$ in (a) at $\phi_\textrm{BL-PL}=120^\circ$ (blue) and 240$^\circ$ (red) with 95\% confidence intervals indicated. (c) Cross section of pumped current along $\phi_{\textrm{BL-PL}}$ at $V_\textrm{PL}=0.823$~V (green dashed line in (a)).}
\label{bothdirections}
\end{figure}

Let us choose the values of $V_\textrm{BR}$, $V_\textrm{BL}$, and $A_\textrm{BR}$ such that they correspond to the maximum plateau length in Fig.~\ref{comparison}(b). Then we measure the pumped current as a function of $V_{\textrm{PL}}$ and $\phi_{\textrm{BL-PL}}$ with the results shown in Fig.~\ref{bothdirections}(a). The pumping plateaus in both directions have the same midpoint, $V_\textrm{PL}=0.823$~V. Thus in comparison to bidirectional pumping observed with two drives in a similar sample\cite{Rossi2014}, we can pump in both directions with fixed $V_\textrm{PL}$ value, whereas with only two drives we needed a different phase and $V_\textrm{PL}$. %In addition pumping is more robust, i.e., the first plateaus in both directions are wider (15 mV and 5 mV in Ref.~\citenum{Rossi2014} and 23 mV and 11 mV here) and we can also perform this pumping with higher frequency (40 MHz in Ref.~\citenum{Rossi2014} and 100 MHz here).

We examine the cross sections along $V_\textrm{PL}$ with two phase differences, $\phi_{\textrm{BL-PL}}=120^\circ$ and $\phi_{\textrm{BL-PL}}=240^\circ$. The absolute value of the pumped current in both cases is shown in Fig.~\ref{bothdirections}(b). In the region where $V_\textrm{PL}=0.820-0.826$~V, the absolute magnitude of current in each direction is the same within the experimental uncertainty. In the middle of this region (at $V_\textrm{PL}=0.823$~V) we study the cross section of Fig.~\ref{bothdirections}(a) along the phase difference $\phi_{\textrm{BL-PL}}$, as shown in Fig.~\ref{bothdirections}(c). We clearly observe that the pumping plateau appears in the negative direction in the range $\phi_{\textrm{BL-PL}}=40-160^\circ$ and in the positive direction in the range $\phi_{\textrm{BL-PL}}=200-290^\circ$.

%We can change the direction of the pumping simply by changing the PL phase only and we do not have to change $V_\textrm{PL}$. This is important since the error counting scheme where we pump electrons back and forth to the reservoir requires  switching between the pumping directions which cannot be achieved swiftly if we need to change the dc voltage of the plunger gate.

\section*{Discussion}

We have demonstrated robust bidirectional pumping with convenient switching between the pumping directions in a silicon single-electron pump with independent rf control over both barriers and the dot potential. Initially, we pump with the entrance barrier and PL, then we introduce a third rf waveform to the exit barrier. We study the effect of the amplitude of the exit barrier swing on the pumped current and observe that it can increase the length of the current plateau in the parameter space of applied gate potentials without the need to increase the amplitudes of the other drives. Subsequently, we studied the pumped current as a function of the plunger gate voltage and the phase difference of the plunger drive with respect to the barrier drives. We find a parameter region where we can perform pumping in positive and negative directions simply by changing the phase of the plunger drive. Furthermore, there is a parameter region where the magnitudes of the pumped currents in different directions are equal within experimental accuracy.

Our architecture, which incorporates individual control over both barriers and the quantum dot is highly flexible. It allows convenient switching between the pumping directions simply by changing the phase of the plunger gate rf drive. This property can potentially be used to extend the basic electron counting scheme demonstrated in Ref.~\citenum{Tanttu2015}, to a more sophisticated error counting protocol\cite{Keller1996}, where single electrons are pumped back and forth between source and reservoir, without accumulating electrons in the reservoir. Different pumping directions may also be used to improve the signal--to--noise ratio in the current measurement by averaging over $+I$ and $-I$ rather than $+I$ and zero. A pumping scheme with three rf drives may also be used to further reduce the uncertainty of output current of the semiconductor quantum dot pumps.

%The fast switching of the pumping direction presented here is important, since it enables the upgrade of the electron counting scheme\cite{Tanttu2015} into an error counting scheme. This scheme allows us to pump single electron back and forth without accumulating electrons to the reservoir. This pumping scheme may also be used to further reduce the uncertainty of the output current of the semiconductor quantum dot pumps.

%We have shown that by introducing a third rf driving voltage to the out-tunneling gate of a silicon quantum dot we can improve the performance of the pump. The length of the plateau is used as an indicator of the robustness of the pumping process. We show that the by increasing the amplitude of the out-tunneling barrier we can increase the length of the plateau in the parameter space. Here we have full rf control of a semiconductor single-electron quantum dot pump and its barriers. In the future, this pumping scheme can potentially be used to further reduce the uncertainty of the output current of the semiconductor quantum dot pumps.

%We also show that it is possible to perform robust bidirectional pumping simply by changing the phase of the rf driving voltage at the plunger gate enables the upgrade of the electron counting scheme\cite{Tanttu2015} into an error counting scheme.

%Some other electron or error counting schemes \cite{ Jehl2003, Kautz2000, Kautz1999, Fujiwara2001, Nishiguchi2006, Yamahata2011, Fricke2013, Fricke2014, Yamahata2014,Yamahata2014_revB} have been presented. 

\section*{Methods}

\subsection*{Sample fabrication}

Our quantum dot is based on a three-layer gate-stack silicon metal--oxide--semiconductor (MOS) technology. The sample is fabricated on a high-purity intrinsic silicon wafer. A 8-nm thick SiO$_2$ layer thermally grown in the active region to form the gate--channel oxide. Three layers of aluminum gates with thicknesses 30, 55, and 80 nm, respectively  are fabricated on top of the wafer with electron beam lithography. Polymethyl methacrylate (PMMA 950k) A4 resist is used as a mask material. After patterning the mask, we develop it in a mixture of methyl isobutyl ketone and isopropanol (MIBK:IPA 1:3) solution. Each layer of Al gates is deposited with a thermal evaporator, followed by a lift-off process. The gates are oxidized on a hot plate in an ambient environment. This process is repeated in total three times to realize all the layers. The source and drain electrodes are connected to the 2DEG with phosphorus-doped regions which form the ohmic contacts. Further details of the fabrication process are discussed elsewhere \cite{Rossi2015}.

%The entire microfabrication and nanofabrication processes used to fabricate this sample and safety measures are explicitly and thoroughly discussed in the literature \cite{Rossi2015}. On top of bulk silicon we have the 8 nm thick layer of SiO$_2$ on top of which we have three layers of aluminum gates.  The three layers of gates are nanofabricated on top of the SiO$_2$ with thicknesses 30, 55, and 80 nm respectively.

\subsection*{Measurement set-up}

All the experiments are performed in a self-made, torlon-based, plastic dilution refrigerator with base temperature of 100 mK submerged in 4-K helium bath. The device is cooled down to base temperature but due to the dissipation on the line impedances, the mixing chamber temperature increases during the experiments up to 800~mK. The device is mounted on a sample holder printed circuit board (PCB) with integrated bias tees with capacitance~10~nF and resistance~100~k$\Omega$. This allows simultaneously application of rf and dc voltages on the driving gates. Our silicon chip is attached to the sample stage with vacuum grease and wedge-bonded to the PCB with Al bond wires. The rf driving voltages from the room temperature set-up are connected to the PCB with coaxial cables with 10 dB attenuation at~4~K. The dc voltages are connected to the PCB with twisted-pair loom lines. The wiring of the sample and the cryostat is discussed in more detail in Ref.~\citenum{Rossi2015}.

A 2DEG is induced at the interface between Si and SiO$_2$ by applying positive voltage to the Al gates. All dc gate voltages are generated by floating dc voltage sources connected to 1:5 voltage dividers. The rf waveforms are generated by an arbitrary waveform generator and synchronized with a rubidium frequency standard. These waveforms are attenuated at room temperature depending on the experiment. The output current is amplified by $10^{10}$~V/A with a transimpedance amplifier powered by a regulated battery pack. The amplified signal is optoisolated to eliminate ground loops and subsequently recorded by a digital multimeter.

%\noindent LaTeX formats citations and references automatically using the bibliography records in your .bib file, which you can edit via the project menu. Use the cite command for an inline citation, e.g.  

\section*{Acknowledgements}
We thank Y. Sun, M. Jenei, G. C. Tettamanzi, I. Iisakka, A.~Kemppinen, J. Lehtinen, and E. Mykk\"anen for fruitful discussions, and M. Meschke and J. Pekola for their design and help in building the cryostat used in the experiments. The financial support from the Centre of Excellence in Computational Nanoscience (project 284621 and 251748) by the Academy of Finland (Grant Nos. 251748, 135794, 272806, and 276528), the Australian Research Council (Grant Nos. DP120104710 and DP160104923), Jenny and Antti Wihuri Foundation, The Finnish Cultural Foundation, and the Australian National Fabrication Facility are acknowledged. A. R. thanks the University of New South Wales Early Career Research Grant scheme for financial support. We acknowledge the provision of facilities and technical support by Aalto University at Micronova Nanofabrication Centre.
%Acknowledgements should be brief, and should not include thanks to anonymous referees and editors, or effusive comments. Grant or contribution numbers may be acknowledged.

\section*{Author contributions statement}

T. T. and A. M. conducted the experiments with support from A. R. and K. Y. T.. T. T. wrote the manuscript with input from all authors. A. R. fabricated the sample. K. W. C. prepared the silicon substrates. A. S. D. and M. M. provided project guidance and supervision.

\section*{Competing financial interests}
The authors declare no competing financial interests.

\section*{Corresponding author}
Correspondence to Tuomo Tanttu
%\section*{Additional information}

%To include, in this order: \textbf{Accession codes} (where applicable); \textbf{Competing financial interests} (mandatory statement). 

%The corresponding author is responsible for submitting a \href{http://www.nature.com/srep/policies/index.html#competing}{competing financial interests statement} on behalf of all authors of the paper. This statement must be included in the submitted article file.

\end{document}